\documentclass[12pt]{article}
\usepackage{latexsym}
\usepackage{float}
\usepackage{amsmath,bm,amssymb}
\usepackage{colordvi}
\usepackage{multicol,multirow}
\usepackage{color}
\usepackage{rotating}
 \usepackage{url,hyperref}
\usepackage[ruled]{algorithm2e}
\usepackage{subfigure}

\usepackage{breqn}
\usepackage{amsfonts}
\usepackage{hyperref}
\usepackage{longtable}

\def\doublespace{\baselineskip=22pt}
\usepackage{lscape}

\newcommand{\overbar}[1]{\mkern 1.5mu\overline{\mkern-1.5mu#1\mkern-1.5mu}\mkern 1.5mu}
\usepackage[super,sort&compress]{natbib}

\begin{document}
\doublespace
\baselineskip 2.8ex
\begin{center}
{\bf\Huge Sparse group variable selection for gene--environment interactions in the longitudinal study  }

\

{\bf \large Fei Zhou$^{1}$, Xi Lu$^{1}$, Jie Ren$^{2}$, Kun Fan$^{1}$, Shuangge Ma$^{3}$ and Cen Wu$^{1\ast}$}
\vspace{0.6em}

{ $^1$ Department of Statistics, Kansas State University, Manhattan, KS}\\

{ $^2$ Department of Biostatistics and Health Data Sciences, Indiana University School of Medicine, Indianapolis, IN}\\

{ $^3$ School of Public Health, Yale University, New Haven, CT}\\

\end{center}

{\bf $\ast$ Corresponding author}:
Cen Wu, wucen@ksu.edu\\
\vspace{0.8em}

\noindent{\bf\Large Abstract}\\

\noindent Penalized variable selection for high dimensional longitudinal data has received much attention as accounting for the correlation among repeated measurements and providing additional and essential information for improved identification and prediction performance. Despite the success, in longitudinal studies the potential of penalization methods is far from fully understood for accommodating structured sparsity. In this article, we develop a sparse group penalization method to conduct the bi-level gene-environment (G$\times$E) interaction study under the repeatedly measured phenotype. Within the quadratic inference function (QIF) framework, the proposed method can achieve simultaneous identification of main and interaction effects on both the group and individual level. Simulation studies have shown that the proposed method outperforms major competitors. In the case study of asthma data from the Childhood Asthma Management Program (CAMP), we conduct G$\times$E study by using high dimensional SNP data as the Genetic factor and the longitudinal trait, forced expiratory volume in one second (FEV1), as phenotype. Our method leads to improved prediction and identification of main and interaction effects with important implications.\\


\section{Introduction}

Longitudinal data have arisen in biomedical studies, clinical trials and many other areas with measurements on the same subject being taken repeatedly over time. Substantial efforts have been made to account for the correlated nature of repeated measures when modelling longitudinal data\cite{VF}. Recently,  the importance of longitudinal design in genetic association studies has been increasingly recognized \cite{Sitlani15,LWLW}. As the main objective of conducting association analysis is to identify key signals associated with the disease phenotypes from a large number of genetic variants (e.g. single nucleotide polymorphisms, or SNPs) \cite{COR},\cite{WuLiCui2012}, the longitudinal design yields novel insight to elucidate the genetic control for complex disease traits over cross-sectional designs.

This study has been partially motivated by analyzing the high dimensional SNP data with longitudinal trait from the Childhood Asthma Management Program (CAMP). CAMP has been launched in early 1990s and became the largest randomized longitudinal clinical trial developed to investigate the long term influences of Budesonide and Nedocromil, the anti-inflammatory therapy, on children with mild to moderate asthma \cite{CAMP1,CAMP2,Covar}. Including placebo, the treatment thus has three levels. Our primary disease phenotype of interest is the forced expiratory volume in one second (FEV1), a repeatedly measured indicator on whether the lung growth of children has improved or not. Here, with SNPs as G factors and treatment, age and gender as environmental (E) factors, we are interested in dissecting the gene-environment (G$\times$E) interactions under the longitudinal trait FEV1. As the number of main and interaction effects is much larger than the sample size, penalized variable selection has become a powerful tool for interaction studies \cite{zhou2021gene}.

To date, penalization methods for interaction studies have been mainly proposed under continuous disease traits, categorical status and cancer prognostic outcomes \cite{zhou2021gene}. With the longitudinal phenotype, where the response on the same subject are repeatedly measured over a set of units (e.g. time), penalized regression methods are relatively underdeveloped for interaction analyses. In fact, our limited literature search indicates that majority of the variable selection methods in longitudinal studies can only accommodate main effects. For example, Wang et al \cite{WZQ} has developed a penalized generalized estimating equation (GEE) for the identification of important main effects associated with longitudinal response. Also within the GEE framework, Ma et al. \cite{MSW} has considered an additive, partially linear model with variable selection on the main effect only. On the other hand, Cho and Qu \cite{CQ} has conducted penalized variable selection in the main effect model based on the quadratic inference function (QIF), and showed that penalized QIF outperforms penalized GEE under a variety of settings. 

The relative underdevelopment of variable selection methods for longitudinal interaction studies is partially due to the challenge in accommodating structured sparsity within either the GEE or QIF framework. Consider the interaction model involving $p$ genetic factors and $q$ environmental factors, where the interactions are denoted by $pq$ product terms. Such a model serves as the umbrella framework for a large number of G$\times$E studies \cite{zhou2021gene}. For one G factor, its main effect and interactions with the $q$ environmental factors form a group of $q$+1 terms. Hence, to determine whether the genetic factor is associated with the phenotype, a group level selection should be conducted. Furthermore, if the genetic factor is associated with the phenotype, an individual level selection within the group is necessary. Overall, identification of important G$\times$E interactions essentially amounts to a sparse group (or bi-level) variable selection problem, which becomes even more challenging when a large number of genetic factors are jointly analyzed under repeatedly measured phenotypes. 

The aforementioned interaction model serves as an umbrella framework for a large number of G$\times$E interaction studies \cite{zhou2021gene}. On a broader scope, sparse group (or bi--level) structure plays a very important role in high dimensional variable selection with structured sparsity \cite{FRI,SimonFriedman2013,BH}. Nevertheless, the bi-level sparsity has not been examined in existing longitudinal studies by far. Our study is novel in that it is among the first to develop the sparse group regularized variable selection for high dimensional longitudinal studies. Specifically, based on the quadratic inference function (QIF), we propose a sparse group variable selection method for simultaneous selection of main and interaction effects on both the group and individual levels in G$\times$E studies. The Minimax concave penalty (MCP) is adopted as the baseline penalty function to achieve regularized identification \cite{Zhang2010}.

Besides the QIF and GEE, Bayesian analysis and mixed models are also the major tools for repeated measurement studies \cite{LWLW,fan2012variable}. Our literature survey shows that the longitudinal bi-level variable selection has not been developed within the two frameworks yet. Therefore, a direct comparison is not possible. While the QIF is robust to model misspecification as well as at least a small portion of data contamination and outliers \cite{qu2004assessing,CQ}, the robustness of Bayesian and mixed model based high dimensional longitudinal analyses remains unanswered. For example, specifying the Bayesian hierarchical model in longitudinal studies generally involves employing a covariance structure, such as the first-order autoregressive (AR1) structure\cite{LWLW}, when the truth is not known $a$ $priori$. It is not clear to what extent these methods are robust to model misspecification. Besides, with the multivariate normal assumption on residual error, Li et al. \cite{LWLW} is not robust to phenotypes with long--tailed distributions. Lastly, we have implemented the proposed and alternative methods in R package \href{https://CRAN.R-project.org/package=springer}{$springer$}     
\cite{SP}. The core modules of the R package have been developed in C++ to guarantee fast computations.

\section{Statistical Methods}
\subsection{Data and Model Settings for Longitudinal G$\times$E Studies}\label{sec:3.1}

We consider a longitudinal scenario where there are $n$ subjects and $k_i$ measurements repeatedly taken over time on the $i$th subject ($1 \leqslant i \leqslant n$). There are correlations among measurements on the same subject, and independence is assumed for measurements between different subjects. We denote $Y_{ij}$ as the phenotypic response of the $i$th subject at the $j$th time point ($1 \leqslant i \leqslant n$, $1 \leqslant j \leqslant k_i$). $X_{ij}=(X_{ij1},...,X_{ijp})^\top$ denotes a $p$-dimensional vector of genetic factors and $E_{ij}=(E_{ij1},...,E_{ijq})^\top$ is a $q$-dimensional vector of environmental factors in the study.  Consider the following model:
\begin{equation}\label{equr:lm}
	\begin{aligned}
		Y_{ij} & =  \mu_{ij}+\epsilon_{ij}\\
		& = \alpha_0 + \sum_{u=1}^{q}\alpha_u E_{iju} + \sum_{v=1}^{p}\gamma_v X_{ijv} + \sum_{v=1}^{p}\sum_{u=1}^{q}h_{uv} E_{iju} X_{ijv}+\epsilon_{ij}\\
		&=\alpha_0 + \sum_{u=1}^{q}\alpha_u E_{iju} + \sum_{v=1}^{p}(\gamma_v + \sum_{u=1}^{q}h_{uv} E_{iju} )X_{ijv}+\epsilon_{ij}\\
		&=\alpha_0 + \sum_{u=1}^{q}\alpha_u E_{iju} + \sum_{v=1}^{p}\eta_v^\top Z_{ijv}+\epsilon_{ij},
	\end{aligned}
\end{equation}
where $\alpha_0$, $\alpha_u$'s, $\gamma_v$'s and $h_{uv}$'s are the coefficients of the intercept, environmental factors, genetic factors and G$\times$E interactions, respectively. We define $\eta_v=(\gamma_v, h_{1v}, ..., h_{qv})^\top$ and $Z_{ijv}=(X_{ijv}, E_{ij1}X_{ijv}, ..., E_{ijq}X_{ijv})^\top$. $Z_{ijv}$ is a $(1+q)$-dimensional vector that represents the main effect of the $v$th genetic factor and its interactions with the $q$ environmental factors. We assume the random error $\epsilon_{i}=(\epsilon_{i1},...,\epsilon_{ik_i})^{\top}\thicksim N_{k_i}(0,\Sigma_{i})$, which is a multivariate normal distribution with $\Sigma_i$ as the covariance matrix for the $k_i$ repeated measurements of the $ith$ subject. Without loss of generality, we let $k_i=k$. Collectively, we can write $\alpha=(\alpha_1, ..., \alpha_q)^\top$, $\eta=(\eta_1^\top, ..., \eta_p^\top)^\top$, and $Z_{ij}=(Z_{ij1}^\top, ..., Z_{ijp}^\top)^\top$.  The vector $\eta$ is of length $p\times(1+q)$. Then model (\ref{equr:lm}) can be rewritten as:
\begin{equation*}
	Y_{ij}=\alpha_0+E_{ij}^\top \alpha+Z_{ij}^\top \eta +\epsilon_{ij}.
\end{equation*}
Denote $ (1+q+p (q+1))$-dimensional vectors $\beta=(\alpha_0,\alpha^\top,\eta^\top)^\top$ and $W_{ij}=(1,E_{ij}^\top,Z_{ij}^\top)^\top$, then model (\ref{equr:lm}) becomes:
\begin{equation*}
	Y_{ij}=W_{ij}^\top \beta +\epsilon_{ij}.
\end{equation*}
While the phenotype, the G factors and E factors all have repeated measurements in the above model formulation for longitudinal G$\times$E studies, such a formulation allows for flexible model setups. For example, it also works when only one of two types of factors is longitudinal, or neither of them have been repeatedly measured. The time-varying gene expression is a representative example of the G factor. In this study, the G factors are SNPs that do not change over time.

\subsection{Quadratic Inference Function for Longitudinal G$\times$E Interactions}\label{sec:3.2}

Modeling longitudinal response $Y_i$ is challenging, as the full likelihood function is generally difficult to specify, due to the intra-subject/cluster correlation. To overcome such an issue, Liang and Zeger\cite{LZ} has proposed the generalized estimating equations (GEE), where a marginal model with only the working correlation for $Y_{ij}$  needs to be specified. The first two marginal moments of $Y_{ij}$ are given as $\text{E}(Y_{ij})=\mu_{ij}=W_{ij}^T\beta$, and $\text{Var}(Y_{ij})=\delta(\mu_{ij})$ respectively, and $\delta(\cdot)$ is a known variance function. The score equation for GEE in the G$\times$E setting is defined as:

\begin{equation*}
	\sum_{i=1}^{n}\frac{\partial \mu_i (\beta)}{\partial \beta}V_i^{-1}(Y_i-\mu_i (\beta))=0,
\end{equation*}
where $\mu_{i} (\beta)=(\mu_{i1}(\beta),...,\mu_{ik}(\beta))^\top$. The first term in the equation, $\frac{\partial \mu_i (\beta)}{\partial \beta}$, reduces to $W_{i}=(W_{i1},...,W_{ik})^\top$. We define $Y_i=(Y_{i1},...,Y_{ik_i})^\top$ and $V_i=A_{i}^{\frac{1}{2}}R_i(\nu)A_{i}^{\frac{1}{2}}$ is the covariance matrix of the $i$th subject, with  $A_{i}=\text{diag}\{\text{Var}(Y_{i1}),...,\text{Var}(Y_{ik})\}$. $R_i(\nu)$ is a `working' correlation matrix that describes the pattern of measurements and can be characterized by a finite dimensional intra--subject/cluster parameter $\nu$.  The solution of the score equation, $\hat{\beta}$, is the GEE estimator.

Liang and Zeger \cite{LZ} has shown that when the intra--subject parameter from the working correlation matrix can be consistently estimated, GEE yields consistent estimates of regression coefficients even if the correlation structure is misspecified. Nevertheless, the GEE estimator is not efficient under such misspecification, let alone the nonexistence of the consistent estimator for the intra--class parameter.  Moreover, the GEE estimator is highly sensitive to even only one outlying observation. To overcome the drawback of GEE, Qu et al. \cite{QLL} has proposed the method of quadratic inference functions (QIF), where a direct estimation of the correlation parameter is not needed, and the corresponding estimator remains optimal even under structure misspecification. In addition, Qu and Song \cite{qu2004assessing} have further shown that QIF is more robust than GEE in the presence of outliers and data contamination, and is thus a preferable method over GEE.

In the current G$\times$E settings, the QIF method approximates the inverse of $R(\nu)$ with a linear combination of basis matrices as
$\text{R}(\nu)^{-1} \approx \sum_{t=1}^{m}b_tM_t$, 
where $M_1$ is an identity matrix, $M_2, ..., M_m$ are  symmetric basis matrices with unknown coefficients $b_1, ... b_m$. Qu et al \cite{QLL} has destribed the choice of the basis matrices $M_2, ..., M_m$ based on the working correlation. With such an approximation, the score equations become

\begin{equation}\label{equr:qifscore}
	\sum_{i=1}^{n}W_{i}^{\top}A_{i}^{-\frac{1}{2}}(b_1M_1+...+b_mM_m)A_{i}^{-\frac{1}{2}}(Y_i-\mu_i (\beta)).
\end{equation}

Within the framework of QIF, we define $\phi_i(\beta)$, the extended score vector involving the main and interaction effects for the $i$th subject, as

\begin{equation}\label{equr:extscore}
	\phi_i(\beta)=\left(\begin{array}{c} W_i^\top A_i^{-\frac{1}{2}}M_1A_i^{-\frac{1}{2}}(Y_i-\mu_i (\beta)) \\ . \\ . \\ . \\
		W_i^\top A_i^{-\frac{1}{2}}M_mA_i^{-\frac{1}{2}}(Y_i-\mu_i (\beta)) \end{array}\right),
\end{equation}
without the estimation of the coefficients $b_1, ... b_m$. Subsequently, the extended score for all subjects is $\overbar{\phi_n}(\beta)=\frac{1}{n}\sum_{i=1}^{n}\phi_i(\beta).$

It can be observed that the estimation functions for G$\times$E studies (Equation (\ref{equr:qifscore})) is equivalent to a linear combination of components from the extended score vectors. Based on $\overbar{\phi_n}(\beta)$, the extended score of the G$\times$E studies, we define the corresponding quadratic inference function as
\begin{equation*}
	Q_n(\beta)=\overbar{\phi_n}^\top(\beta)\overbar{\Omega_n}(\beta)^{-1}\overbar{\phi_n}(\beta), 
\end{equation*} 
where $\overbar{\Omega_n}(\beta)=\frac{1}{n}\sum_{i=1}^{n}\phi_i(\beta)\phi_i(\beta)^\top$. Then the QIF estimator $\hat{\beta}$ for G$\times$E interaction studies can be obtained as $\hat{\beta}=\text{arg}\underset{\beta}{\text{min}}Q_n(\beta)$.

\subsection{Penalized identification of  G$\times$E interactions in longitudinal studies}\label{sec:3.3}

In a typical G$\times$E study, the main objective is to identify an important subset of features out of all the main and interaction effects, which is of a ``large $p$, small $n$'' nature. Therefore, penalized variable selection becomes a natural tool to investigate G$\times$E interactions \cite{zhou2021gene}. With model (\ref{equr:lm}), we propose the following penalized quadratic inference function:
\begin{equation}\label{equr:sgl}
	U(\beta)=Q(\beta)+\sum_{v=1}^{p}\rho(||\eta_{v}||_{\Sigma_{v}};\lambda_{1},\gamma)+\sum_{v=1}^{p} \sum_{u=1}^{q+1}\rho  (|\eta_{vu}|;\lambda_{2},\gamma),
\end{equation}
\noindent where the baseline penalty function $\rho(\cdot)$ is a minimax concave penalty, which is defined as $\rho(t;\lambda,\gamma)=\lambda \int_{0}^{t}(1-\frac{x}{\gamma \lambda})_{+}dx$ on $[0,\infty)$, with tuning parameter $\lambda$ and regularization parameter $\gamma$ \cite{Zhang2010}. As previously defined, $\eta_v$ is a coefficient vector of length $q+1$, corresponding to the main effect of the $v$th SNP and its interactions with the $q$ environment factors. We denote $||\eta_{v}||_{\Sigma_{v}}$ as the empirical norm of $\eta_{v}$ and $\eta_{vu}$ as the $u$th component of $\eta_v$($v=1,...,p, \text{and } u=1,...,q+1$).  

Our choice of the baseline penalty function is the minimax concave penalty and the corresponding first derivative function of MCP penalty is defined as $\rho' (t;\lambda,\gamma)=(\lambda-\frac{t}{\gamma})I(t \leq \gamma \lambda)$. 

Within the current longitudinal setting, identification of important G$\times$E interactions amounts to a bi--level selection problem. In particular, selection on the group level determines whether the genetic factor is associated with the phenotypic response. If the coefficient vector $\eta_{v}$ is 0, then the G factor does not have any contribution to the response. Otherwise, an examination on the individual level to further determine the existence of main and interaction effects is necessary. The penalized QIF function (\ref{equr:sgl}) has been formulated to accommodate individual and group level selection in longitudinal G$\times$E studies with the sparse group MCP penalty function. 

In general, the regularized loss functions of penalization problems share the form of ``unregularized loss function + penalty function'' \cite{WuMa2015}. In longitudinal studies, popular choices of unregularized loss function include GEE and QIF. Our limited search suggests that existing penalization methods for longitudinal data are mostly focused on main effects, therefore only baseline penalty functions such as LASSO and SCAD are necessary \cite{WZQ,MSW,CQ}. In G$\times$E studies, the interaction structure poses a challenge to accommodate the more complicated bi-level sparsity, which has motivated the proposed study.

\subsection{Computational Algorithms for Sparse Group QIF}\label{sec:3.4}

Now, we outline an efficient Newton-Raphson algorithm that iteratively updates parameter estimates $\hat{\beta}$ for the penalized QIF. In particular, at the $g$th iteration, $\hat{\beta}^{(g+1)}$ can be obtained based on the estimated coefficient vector $\hat{\beta}^{(g)}$ from the $g$th iteration as follows:
\begin{equation}\label{equr:2}
	\hat{\beta}^{(g+1)}=\hat{\beta}^{(g)}+[V^{(g)}+nH^{(g)})]^{-1}[P^{(g)}-nH^{(g)}\hat{\beta}^{(g)}],
\end{equation}
where $P^{(g)}$ and $V^{(g)}$ are the first and second order derivative functions of the score function of QIF, respectively. They are given as:
\begin{equation*}
	P^{(g)}=\frac{\partial Q(\hat{\beta}^{(g)})}{\partial \beta}=2\frac{\partial \overbar{\phi_n}^\top}{\partial \beta} \overbar{\Omega_n}^{-1}\overbar{\phi_n}(\hat{\beta}^{(g)}),
\end{equation*}
and 
\begin{equation*}
	V^{(g)}=\frac{\partial^2 Q(\hat{\beta}^{(g)})}{\partial^2 \beta}=2\frac{\partial \overbar{\phi_n}^\top}{\partial \beta} \overbar{\Omega_n}^{-1}\frac{\partial \overbar{\phi_n}}{\partial \beta}.
\end{equation*}
Besides, $H^{(g)}$ is a diagonal matrix consisting of derivatives of both the individual-- and group-- level penalty functions, which is defined as:
\begin{equation*}
	\begin{aligned}
		H^{(g)}&=\text{diag}(\underbrace{0,...,0}_{1+q},\underbrace{\frac{\rho'(||\hat{\eta}_{1}^{(g)}||_{\Sigma_{1}};\sqrt{q+1}\lambda_{1},\gamma)}{\epsilon+||\hat{\eta}_{1}^{(g)}||_{\Sigma_{1}}},...,\frac{\rho'(||\hat{\eta}_{1}^{(g)}||_{\Sigma_{1}};\sqrt{q+1}\lambda_{1},\gamma)}{\epsilon+||\hat{\eta}_{1}^{(g)}||_{\Sigma_{1}}}}_{1+q},...,\\ &\underbrace{\frac{\rho'(||\hat{\eta}_{p}^{(g)}||_{\Sigma_{p}};\sqrt{q+1}\lambda_{1},\gamma)}{\epsilon+||\hat{\eta}_{p}^{(g)}||_{\Sigma_{p}}},...,\frac{\rho'(||\hat{\eta}_{p}^{(g)}||_{\Sigma_{p}};\sqrt{q+1}\lambda_{1},\gamma)}{\epsilon+||\hat{\eta}_{p}^{(g)}||_{\Sigma_{p}}}}_{1+q})+\text{diag}(\underbrace{0,...,0}_{1+q},\\ &\underbrace{\frac{\rho'(|\hat{\eta}_{11}^{(g)}|;\lambda_{2},\gamma)}{\epsilon+|\hat{\eta}_{11}^{(g)}|},...,\frac{\rho'(|\hat{\eta}_{1(q+1)}^{(g)}|;\lambda_{2},\gamma)}{\epsilon+|\hat{\eta}_{1(q+1)}^{(g)}|}}_{1+q},...,\underbrace{\frac{\rho'(|\hat{\eta}_{p1}^{(g)}|;\lambda_{2},\gamma)}{\epsilon+|\hat{\eta}_{p1}^{(g)}|},...,\frac{\rho'(|\hat{\eta}_{p(q+1)}^{(g)}|;\lambda_{2},\gamma)}{\epsilon+|\hat{\eta}_{p(q+1)}^{(g)}|}}_{1+q}),
	\end{aligned}
\end{equation*}
where $\epsilon$ is a small positive number adopted to ensure that the denominator is nonzero for zero coefficients and here we set it equal to $10^{-6}$. This is a common practice to avoid numerical instability in Newton--Raphson type of algorithms. The first $(1 + q)$ elements on the diagonal of matrix $H^{(g)}$ are zero, which indicates no shrinkage is added to the intercept and the coefficients of the environmental factors. Here $nH^{(g)}\hat{\beta}^{(g)}$ and $nH^{(g)}$ can be used to approximate the first and second order derivative functions of the sparse group penalty, respectively. Given an initial coefficient vector, which can be estimated by LASSO, the proposed algorithm proceeds iteratively and update the regression parameter $\hat{\beta}^{(g+1)}$ until convergence which can be achieved when the L1 norm of the difference in coefficient vectors from adjacent iterations is less than 0.001. Our numerical experiments show that the convergence can usually be achieved in a small to moderate number of iterations. 

There are usually two tuning parameters for sparse group penalty, controlling the individual and group level sparsity, respectively. In the current G$\times$E study, for a G factor, its main effect and interactions with all the environmental factors are treated as one group. The tuning parameter $\lambda_1$ determines the amount of shrinkage on the group level, and $\lambda_2$ further tunes the shrinkage on  individual effects within the group. The optimal pair of $\lambda_1$ and $\lambda_2$ are obtained through a joint search over a two-dimensional grid of ($\lambda_1, \lambda_2$) based on a validation approach. Specifically, the regularized estimate is computed on a training dataset, and then the prediction is evaluated on an independently generated testing dataset. Our numerical experiment shows that validation and cross validation tend to yield similar tunings, but the first one is computationally much faster. 

With the nonconvex baseline penalty MCP, we will need to determine the regularization parameter $\gamma $ which balances unbiasedness and concavity \cite{Zhang2010}. Relevant studies suggests checking with a sequence of different values, and then fixing the value. We have investigated a sequence of 1.4, 3, 4.2, 5.8, 6.9, and 10, and found that the results are not sensitive to the value of $\gamma $. Therefore, we set $\gamma $ to 3. This finding is consistent with published studies \cite{Ren17, WU18}.   

For fixed tuning parameters, the proposed algorithm proceeds as follows:

(a) Initialize the coefficient vector $\hat{\beta}^{(0)}$ using LASSO;

(b) At the $(g+1)$th iteration, update $\hat{\beta}^{(g+1)}$ based on equation (\ref{equr:2}) ;

(c) Repeat Step (b) until the convergence is achieved.

We consider three working correlation structures, exchangeable, AR(1) and independence, for the sparse group MCP based method dissecting longitudinal G$\times$E interactions. Besides, the group MCP which ignores the within group sparsity of G$\times$E interactions and the MCP only considering individual level main and interaction effects are included for comparison. In summary, we term the bi--level, group--level and individual--level longitudinal penalization under exchangeable working correlation as sgQIF.exch, gQIF.exch and iQIF.exch, respectively. Similarly, with AR(1) correlation, the three approaches are denoted as sgQIF.ar1, gQIF.ar1 and iQIF.ar1 correspondingly. Then sgQIF.ind, gQIF.ind, and gQIF.ind are termed accordingly under independent correlation. The details of the alternative approaches are provided in Appendix \ref{appendix:b}. We computed the CPU running time for 100 replicates of simulated gene expression data with n = 400, $p$ = 200, $q$=5
(with a total dimension of 1206) and fixed tuning parameters on a regular laptop for the nine methods, which can be
implemented using our developed package: $springer$ \cite{SP}. 
The average CPU running time in seconds are 34.7 (sd 4.9) (sgQIF.exch), 36.2 (sd 6.9) (gQIF.exch), 35.7 (sd 3.5) (iQIF.exch), 24.9 (sd 4.3) (sgQIF.ar1), 32.7 (sd 1.5) (gQIF.ar1), 26.5 (sd 5.3) (iQIF.ar1), 5.8 (sd 0.5)(sgQIF.ind), 6.3 (sd 0.8) (gQIF.ind) and 5.4 (sd 0.3) (iQIF.ind), respectively.

\subsection{Unbalanced Date Implementation}\label{sec:3.5}

In practice, due to missing data, the repeated measurements are unbalanced when cluster sizes vary among different subjects. The proposed method can still be implemented in such a case by introducing a transformation matrix to each subject\cite{CQ}.  Suppose the total number of time points is denoted by $k$ and the $i$th subject is repeated measured at $k_i$ time points.  Let $S_i$ denote a $k\times k_i$ tranformation matrix for the $i$th subject.  Then for the $i$th subject, the transformation matrix $S_i$ is generated by deleting the columns of the $k \times k$ identity matrix that correspond to the time points with measurement missing.  According to this strategy, transformation is performed by letting $W_i^{\star}=S_i W_i, Y_i^{\star}=S_i Y_i, \mu_i^{\star}(\beta)=S_i \mu_i(\beta)$ and $A_i^{\star}=S_i A_i S_i^{\top}$.  Then we can replace $\phi_i(\beta)$ in equation (\ref{equr:extscore}) by the transformed extended score vector $\phi_i^{\star}(\beta)$, which is defined as:

\begin{equation*}
	\phi_i^{\star}(\beta)=\left(\begin{array}{c} (W_i^{\star})^\top (A_i^{\star})^{-\frac{1}{2}}M_1(A_i^{\star})^{-\frac{1}{2}}(Y_i^{\star}-\mu_i ^{\star}(\beta)) \\ . \\ . \\ . \\
		 (W_i^{\star})^\top (A_i^{\star})^{-\frac{1}{2}}M_m(A_i^{\star})^{-\frac{1}{2}}(Y_i^{\star}-\mu_i ^{\star}(\beta)) \end{array}\right),
\end{equation*}
and  the QIF estimator can be further obtained for unbalanced data based on the transformed terms.

\section{Simulation}

The performance of the nine methods has been assessed through simulation studies to demonstrate the utility of the proposed methods.  We generate the responses from model (\ref{equr:lm}) with sample size $n$=400, and set the number of time points $k$ to 5. The dimension for genetic factors is $p$= 200 and there are $q=5$ environmental factors.  This leads to a total dimension for all the main and interaction effects equal to 1206, which is much larger than the sample size. We have also experimented with larger dimensionality for the G factors, and found that the results are stable and consistent with the current setting as long as the total dimensionality is moderately larger than sample size. The details on scalability of the proposed method to ultra-high dimensional data is deferred to the Section of Discussion. In our simulation, the environmental factors are simulated from a multivariate normal distribution with mean 0 and AR--1 covariance matrix with marginal variance 1 and auto correlation coefficient 0.8. The first environmental factor is dichotomized at the 50th percentile and changed to a binary vector. We simulate the random error $\epsilon$ for the longitudinal response from a multivariate normal distribution by assuming 0 mean vector and an exchangeable covariance structure with parameter $\tau=0.8$. Following all these settings, the time-independent genetic factors are simulated in four different scenarios.

In the first scenario, the genetic factors are gene expressions, which are simulated from a multivariate normal distribution with mean 0 and AR--1 covariance matrix with marginal variance 1 and an auto correlation coefficient 0.8.  In the second scenario, we consider generating SNP data by dichotomizing the gene expression values from scenario 1 at the 30th and 70th percentiles with respect to each gene, leading to the three categories (0,1,2) for genotypes (aa,Aa,AA). 

In the third scenario, we simulate the SNP data using a pairwise linkage disequilibrium (LD) structure. Let $q_A$ and $q_B$ denote the minor allele frequencies (MAFs) for the two risk alleles A and B from two adjacent SNPs, respectively, and $\delta$ denote the LD.  Then the frequencies of the four haplotypes can be derived as $p_{AB}=q_Aq_B+\delta, p_{ab}=(1-q_A)(1-q_B)+\delta, p_{Ab}=q_A(1-q_B)-\delta,$ and $p_{aB}=(1-q_A)q_B-\delta$.  By assuming Hardy-Weinberg equilibrium, we simulate the SNP genotypes AA, Aa and aa at locus 1 from a multinomial distribution with frequencies $q_A^2, 2q_A(1-q_A)$ and $(1-q_A)^2$.  Then the genotypes for SNP at locus 2 can be generated based on the conditional genotype probability matrix \cite{Cui}.  If the MAFs are 0.3 and pairwise correlation r is set to 0.3, we can get $\delta=r\sqrt{q_A(1-q_A)q_B(1-q_B)}$.

Next, in scenario 4, we consider a more practical approach to generate the SNP data. The first 200 SNPs from the case study have been extracted as the G factors.  We randomly sample 400 subjects from the real data in each simulation replicate to generate the longitudinal responses.

The coefficients are generated from Uniform[0.3, 0.7] for 31 nonzero effects, consisting of the intercept, 5 environmental factors, and 25 genetic main effects and G$\times$E interactions. We simulate 100 replicates for each scenario to evaluate the identification and prediction performance of all the 9 methods. The average number of true positives (TP) and false positives (FP) with the corresponding standard deviation (sd) are recorded. In addition, prediction accuracy is evaluated based on the mean squared error.

Identification results are tabulated in Tables ~\ref{tab:1},~\ref{tab:2} in the main text, and Tables ~\ref{tab:3} and \ref{tab:4} in Appendix \ref{appendix:c}.  In general, the proposed sparse group G$\times$E interactions under the exchangeable(sgQIF.exch), AR1(sgQIF.ar1) and independence (sgQIF.ind) working correlation structures outperform the alternatives focusing only on the group level effects (gQIF.exch, gQIF.ar1 and gQIF.ind) and individual level effects (iQIF.exch, iQIF.ar1 and iQIF.ind). Table ~\ref{tab:1} shows the result of using gene expressions as G factors from the first scenario with $n=400, p=200, \tau=0.8$. There are 25 important main and interaction effects with corresponding nonzero coefficients. Under the exchangeable working correlation, sgQIF.exch identifies 21.4 (sd 1.1) true positives, while the number of false positives, 2.6 (sd 1.5), is relatively small. On the other hand, iQIF.exch only considers the individual main and interaction effects, yielding 21.6 (sd 1.1) true positives, with 6.4 (sd 5.2) false positives. gQIF.exch identifies an FP of 14.8 (5), which is the largest number of false positives among the three under the same working correlation structure. It is also worth noting that the standard deviations associated with the alternative approaches, i.e. 5 for gQIF.exch and 5.2 for iQIF.exch, are quite larger than that of the proposed one (1.5 for sgQIF.exch). A closer look over the results shows that such all these differences mainly come from the identification of interaction effects. sgQIF.exch has the smallest FP (2.4  with sd 1.3) for the interaction effects, followed by iQIF.exch (5.4 with sd 4.6) and gQIF.exch (14.4 with sd 4.5).

\begin{table} [ht!]
	\begin{center}
		\caption{Identification results for Scenario 1. TP/FP: true/false positives.  mean(sd) of  TP and FP based on 100 replicates.}\label{tab:1}
		
		\begin{tabular}{ l c c c c c c  }
			\hline
			&\multicolumn{2}{c}{Overall }&\multicolumn{2}{c}{Main } &\multicolumn{2}{c}{Interaction } \\
			\cline{2-7}
			&TP&FP&TP&FP&TP&FP\\
			\hline
			sgQIF.exch&21.4(1.1)&2.6(1.5)&5.4(1.1)&0.2(0.4)&16.0(1.9)&2.4(1.3)\\
			
			gQIF.exch&23.4(1.1)&14.8(5.0)&6.0(1.2)&0.4(0.9)&17.4(0.9)&14.4(4.5)\\
			
			iQIF.exch&21.6(1.1)&6.4(5.2)&5.4(1.1)&1.0(1.7)&16.2(1.9)&5.4(4.6)\\
			
			\hline
			sgQIF.ar1&21.7(1.2)&3.2(1.9)&5.5(1.0)&0.3(0.5)&16.2(1.7)&2.8(1.6)\\
			
			gQIF.ar1&23.7(1.2)&14.8(4.4)&6.2(1.2)&0.3(0.8)&17.5(0.8)&14.5(4)\\
			
			iQIF.ar1&21.8(1.2)&6.2(4.7)&5.5(1.0)&1.0(1.5)&16.3(1.8)&5.2(4.1)\\
			
			\hline
			sgQIF.ind&20.7(1.0)&2.7(2.2)&4.5(1.2)&0.2(0.4)&16.2(0.8)&2.5(1.9)\\
			
			gQIF.ind&22.3(1.2)&16.5(7.0)&5.5(1.0)&1.0(1.5)&16.8(0.8)&15.5(5.5)\\
			
			iQIF.ind&21.0(0.9)&5.2(3.1)&4.5(1.2)&0.5(0.8)&16.5(0.8)&4.7(2.3)\\
			\hline 
		\end{tabular}
	\end{center}
	\centering
\end{table}

\begin{table} [ht!]
	\begin{center}
		\caption{Identification results for Scenario 2. TP/FP: true/false positives.  mean(sd) of  TP and FP based on 100 replicates.}\label{tab:2}
		
		\begin{tabular}{ l c c c c c c }
			\hline
			&\multicolumn{2}{c}{Overall }&\multicolumn{2}{c}{Main } &\multicolumn{2}{c}{Interaction } \\
			\cline{2-7}
			&TP&FP&TP&FP&TP&FP\\
			\hline
			sgQIF.exch&19.4(0.7)&1.3(1.2)&3.3(0.7)&0.1(0.4)&16.1(0.6)&1.1(1.0)\\
			
			gQIF.exch&21.5(1.9)&13.3(4.0)&4.4(1.6)&0.5(0.8)&17.1(0.6)&12.8(3.3)\\
			
			iQIF.exch&20.1(1.2)&4.4(4.0)&3.3(1.0)&0.5(0.8)&16.9(0.6)&3.9(3.4)\\
			
			\hline
			
			sgQIF.ar1&19.0(0.9)&1.0(1.0)&3.3(0.6)&0.1(0.4)&15.7(0.6)&1.0(1.0)\\
			
			gQIF.ar1&21.7(2.9)&12.7(4.0)&4.7(2.1)&0.3(0.6)&17.0(1.0)&12.3(3.5)\\
			
			iQIF.ar1&20.7(0.6)&6.7(5.7)&3.7(0.6)&0.7(1.2)&17.0(1.0)&6.0(4.6)\\
			
			\hline
			
			sgQIF.ind&19.0(2.0)&1.8(0.7)&3.3(1.2)&0.1(0.4)&15.7(1.2)&1.6(0.7)\\
			
			gQIF.ind&21.3(1.3)&15.5(8.2)&3.8(0.9)&0.8(1.8)&16.5(0.8)&14.8(7.0)\\
			
			iQIF.ind&19.5(1.8)&5.3(3.6)&3.5(0.9)&1.1(1.2)&16.0(1.3)&4.1(3.0)\\
			\hline 
		\end{tabular}
	\end{center}
	\centering
\end{table}

\begin{figure}[ht!]
	\caption{Identification results under 25 important genetic main effects and G$\times$E interactions (corresponding to 25 nonzero regression coefficients) in the 4 scenarios. TP/FP: true/false positives.  mean(sd) of  TP and FP based on 100 replicates.}
	\includegraphics[width=1\textwidth]{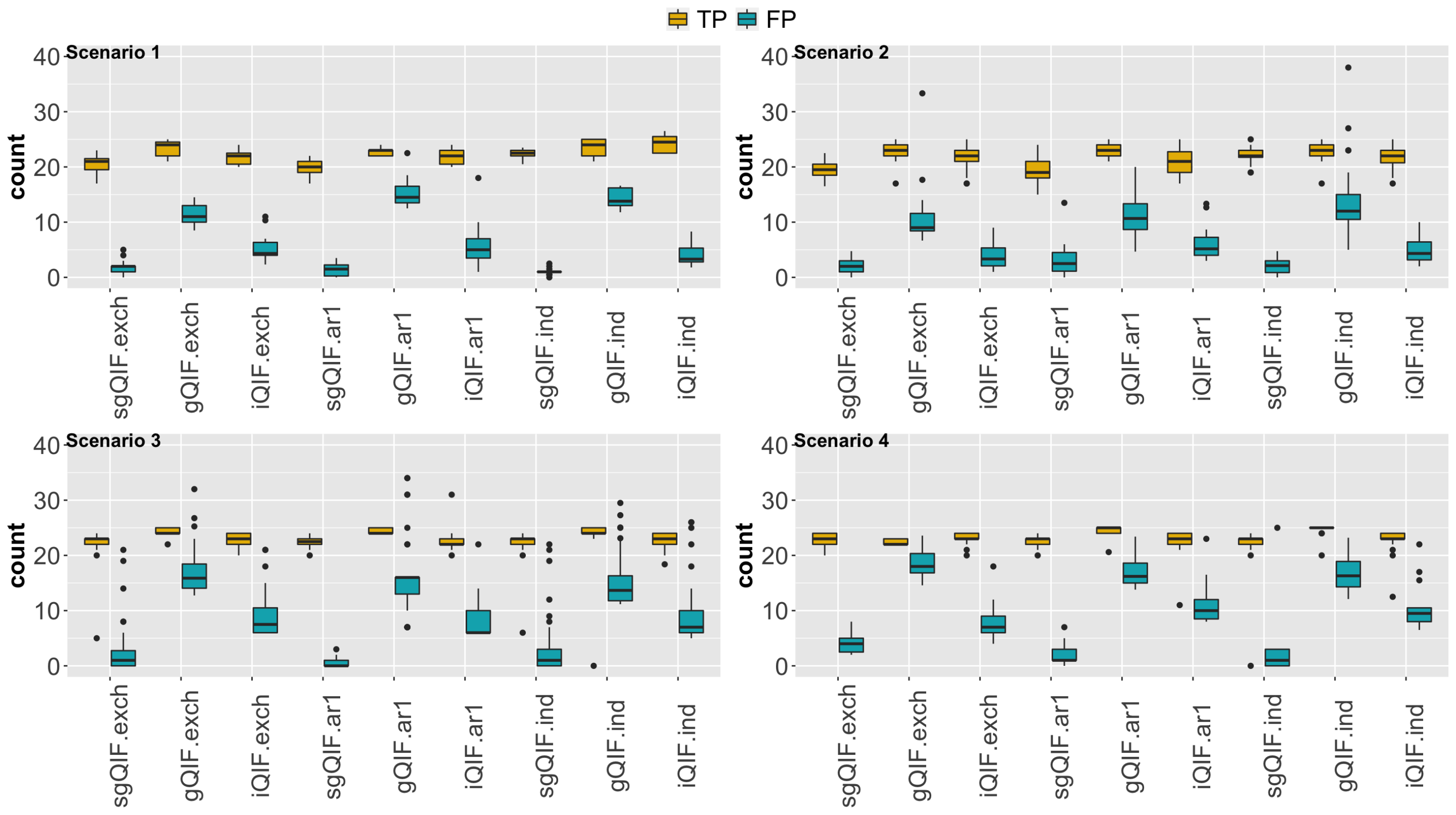}
	\label{fig:1}
\end{figure}

Similar patterns can be observed from other settings. For instance, Table ~\ref{tab:2} displays the result for the simulated SNP data from Scenario 2.  sgQIF.exch identifies an TP of 19.4 (sd 0.7) with 1.3 (sd 1.2) false positives.  gQIF.exch has 21.5 (sd 1.9) true positives with a much larger number of false positives 13.3 (sd 4.0). The number of TP and FP pinpointed by iQIF.exch are 20.1 (sd 1.2) and 4.4 (sd 4.0), respectively. Under the same exchangeable working correlation, while the number of identified TPs are comparable, both the average and standard deviations of alternatives are much larger than the proposed method. The identification results for the 4 scenarios are displayed in Figure \ref{fig:1}, which clearly shows that the proposed method outperforms the competing alternatives in the identification of longitudinal G$\times$E interactions. Figure \ref{fig:2} summarizes the prediction results of the 4 scenarios.  In Scenario 1 under the exchangeable working correlation, sgQIF.exch has a prediction error less than that of the gQIF.exch and iQIF.exch. We have similar findings in other settings as well, which indicates the proposed bi-level method has superior prediction performance over the group level and individual level based methods.

In longitudinal studies, the QIF framework is robust to the misspecification of working correlations \cite{QLL}. In our simulation, although the results without misspecifying working correlation appear to be better, overall, they are comparable across different settings. Such a property is especially appealing when the ground truth on working correlation is not available. Another fold of robustness in QIF comes from its insensitivity to small portions of outlying observations and data contamination, which has been theoretically and empirically investigated in Qu and Song \cite{qu2004assessing}. Meanwhile, the GEE based ones, as well as models assuming Gaussian responses and working independence among repeated outcomes, are not robust and lead to biased results given the presence of even a single outlier. A comprehensive evaluation of this fold of robustness is beyond the scope this study, and will be conducted in the near future.

\section{Real Data Analysis}

Asthma is a chronic respiratory disease with lung inflammation and reversible airflow obstruction, resulting in difficulty in breath. According to the Centers for Disease Control and Prevention (CDC), more than 25 million Americans have asthma.  7.7 percent of adults and 8.4 percent of children in the U.S. have asthma \cite{CDC1,CDC2}. Asthma is the leading chronic disease among children. We analyze the data from Childhood Asthma Management Program (CAMP) in our case study \cite{CAMP1,CAMP2,Covar}. The SNP and phenotype datasets (with accession pht000701.v1.p1) from CAMP have been downloaded and pre--processed. Subjects who are 5 to 12 years and diagnosed with chronic asthma have been selected and monitored over 4 years. There are three visits before treatment with each visit 1-2 weeks apart.  Thirteen visits are made after treatment. The first two visits after treatment are 2 months apart and the remaining visits are 4 months apart. The twelve visits that are 4 months apart after treatment are selected in our study. Two types of treatments are given to the subjects. Treatments Budesonide and Nedocromil are assigned to 30\% of the subjects, and the rest receive placebo. We consider treatment, age and gender as environmental factors. The phenotype of interest is the forced expiratory volume in one second (FEV1), which is the total volume of air expelled out of the lung in one second and it's repeatedly measured during each visit.  The genotype information of each subject contains over nine hundred thousand SNPs.  We match genotypes with phenotypes based on subject id's and remove the SNPs with minor allele frequency (MAF) less than 0.05 or deviation from Hardy-Weinberg equilibrium and obtain a working dataset with 438 Caucasian subjects and 447,850 SNPs. 

For computational convenience in studies with ultrahigh dimensionality, such as the Genome Wide Association Studies (GWAS) and multi--omics integration studies, marginal feature prescreening needs to be conducted first so that regularization can be applied on datasets with reasonably large scale \cite{fan2010selective,wu2019selective}. For instance,  Li et al.\cite{LWLW}, Jiang et al. \cite{jiang20152higwas}  and Wu et al.\cite{WCM} have adopted single SNP analysis for prescreening before applying the proposed variable selection methods in longitudinal and multivariate GWAS studies, respectively. Here, we use a marginal G$\times$E model with FEV1 as the response to filter SNPs. The predictors of the marginal model consist of E factors, the single SNP main effect, as well as their interactions. The SNPs with at least one of the p values that correspond to G and G$\times$E interactions in the marginal model less than a certain cutoff (0.005) are kept. 261 SNPs have passed the screening.

We apply the method sgQIF.exch under the exchangeable working correlation and analyze the data together with the alternative method iQIF.exch, which consider all the effects individually.  The optimal tuning parameters are achieved through a 5-fold cross-validation. We obtain the predicted mean squared error after refitting using selected variables from the orginal data. sgQIF.exch has a smaller prediction error (0.16) than that of iQIF.ind (0.23). The identification results are tabulated in tables ~\ref{tab:5} and \ref{tab:6} in Appendix \ref{appendix:a}. Methods that consider group effects only show inferior performance in the simulation study are not adopted in the real data analysis.  The proposed method sgQIF.exch identifies 130 effects in total,  34 of which are genetic main effects and the remaining 96 are interaction effects.  iQIF.exch totally identifies 130 effects, with 28 genetic main effects and 102 interaction effects.  sgQIF.exch and iQIF.exch commonly identify 22 genetic main effects and 62 interaction effects.  There are twelve SNPs that are uniquely identified by the proposed method sgQIF.exch and they will provide some useful implications. They can be mapped to the corresponding genes and some of the genes have been found to be related to the development of asthma.  For instance, sgQIF.exch identifies the main effect of the SNP rs17390967 and its interactions with the environment factors treatment and gender.  The SNP rs17390967 is located within the gene SCARA5.  SCARA5 is a member of the scavenger receptor A (SR-A), which is found to be protective to the lung using the ovalbumin-asthma model of lung injury \cite{Arredouani}. The interaction with treatment indicates that the expression level of SCARA5 may influence the effect of medical therapy in the treatment of asthma. Another example is the SNP rs767006, which is located in the gene CYFIP2. The prposed method sgQIF.exch identifies the main effect of rs767006 and its interaction with gender. CYFIP2, together with CYFIP1 make up the CYFIP family.  It has been found that there is a strong association between asthma and polymorphisms in CYFIP2 \cite{Noguchi}. Method sgQIF.exch also identifies rs6914953 and its interaction with gender.  The identified SNP rs6914953 is located in F13A1.  F13A1 codes for the $\alpha$ subunit of Factor X111, which is the last enzyme generated in the blood coagulation cascade and it stabilizes blood clots with cross-linking fibrin. F13A1 has been considered as a susceptible locus for obesity and it has been found that there is a consistent link between asthma and obesity \cite{Sharma}.  Another identified SNP is rs4647108, that is mapped to the gene ERCC8. ERCC8 has also been found to be related with the development of asthma \cite{Wilson}. The method sgQIF.exch identifies the main effect of rs4647108 and its interaction with gender.  This result is consistent with previous findings that over-expression of ERCC8 is associated with a higher FEV1, which indicates a development of asthma.

\section{Discussion}

In general, regularization methods work well when the dimensionality is up to the order that is moderately larger than sample size. To handle ultra-high dimensional data, the two stage variable selection consisting of a quick marginal screening stage and a post-screening refining stage with the direct applications of regularization has been widely used \cite{fan2010selective}, including the longitudinal GWAS \cite{LWLW,jiang20152higwas}. The marginal feature screening, preferably with theoretical guarantees such as the sure independence screening \cite{fan2008sure,song2014censored,li2014fast},  is necessary for reducing the ultra-high dimensionality of features to a reasonable order so regularized variable selection is applicable  \cite{fan2010selective}. By far, consensus on the optimal screening strategy with repeated measurements has not been reached yet. In this study, we have adopted a marginal G$\times$E model to conduct screening, which is more consistent with the nature of regularization at the refining stage.

There are published studies on variable selection in varying coefficient models with repeated measurements (Wang et al. \cite{WANG08}, Noh and Park \cite{NOH10} and Tang et al. \cite{Tang13}, among others). A common limitation in these studies is that the within--subject correlation has not been taken into account. From the perspective of G$\times$E interactions, the time varying effects investigated in these studies can be viewed as nonlinear G$\times$E interactions \cite{MA11,WUC,WZC,wu2015penalized,li2020semiparametric,ma2015semiparametric}. In our study, the interaction effects is modeled as the product between G and E factors, which is under the linear G$\times$E interaction assumption \cite{zhou2021gene}. To the best of our knowledge, no published studies have been developed for variable selection in G$\times$E interaction studies with linear assumptions.

The bi--level structure plays a critical role in studies concerning the more general linear G$\times$E interactions \cite{zhou2021gene}. The key contribution of the proposed study lies in developing sparse group regularization within the QIF framework to accommodate within--cluster correlations among repeated measurements. As a major competitor of GEE, QIF is more efficient when the working correlation is misspecified. Our work is significantly different from Zhou et al.\cite{ZHOU19} in that the lipid--environment interaction analysis of repeated measurements has been developed based on GEE, and, more importantly, the interaction is pursued only on a group level and does not lead to the within group sparsity. So it is not applicable to the current setting.

This study can be extended in multiple horizons. For instance, marginal regularization has been demonstrated as an effective approach to dissect G$\times$E interactions \cite{ZHANG20,lu2021identifying}. Our methods can be readily adopted to conduct marginal identification of interaction effects when the phenotypes are repeatedly measured. In addition, robust variable selection for G$\times$E interactions have been proposed  \cite{WJRCM,REN20,zhang2020robust}. In longitudinal G$\times$E studies, the robustness of QIF framework to data contamination in the response can be potentially improved by modifying the weight in estimating equation to downweigh the influences of outliers. Recently,  Wang et al.\cite{wang2021modeling} have revealed the benefit of accounting for network structure in large scale G$\times$E studies. By incorporating the network constrained regularization, the proposed method can better accomodate the correlation among SNPs due to linkage disequilibrium.

\bibliography{references}

\newpage
\clearpage
\section*{Appendix}
\appendix
\section{Derivations of Alternative Methods}
\label{appendix:b}

The alternative methods fall into the following two categories: (1) gQIF.exch, gQIF.ar1 and gQIF.ind only conduct penalized identification on the group level, corresponding to the penalized group QIF,  and (2) iQIF.exch, iQIF.ar1 and iQIF.ind ignore the group level effects, and only focus on the individual level effects (penalized QIF).

\subsection{Penalized Group QIF}

The penalized group QIF methods considered in this study  (gQIF.exch, gQIF.ar1 and gQIF.ind) can only identify the main and interaction effects on a group--in/group--out basis. The corresponding score equation is defined as
\begin{equation*}
	U(\beta)=Q(\beta)+\sum_{v=1}^{p}\rho(||\eta_{v}||_{\Sigma_{v}};\sqrt{q+1}\lambda_{1},\gamma),
\end{equation*}
where $\rho$ denotes MCP penalty with tuning parameter $\lambda_{1}$ and regularization parameter $\gamma$. As defined in Section \ref{sec:3.2}, the coefficient vector $\beta$ corresponds to all the main and interaction effects. $\eta_v$, the vector of length $q$+1 in $\beta$, represents the main effect of the $v$th G factor as well as its interactions with the $q$ environment factors. The penalty is imposed on $||\eta_{v}||_{\Sigma_{v}}$, the empirical norm of $\eta_{v}$. Thus the penalized identification can merely performed on group level. 

We have developed a Newton-Raphson based algorithm to obtain the penalized QIF estimate $\hat{\beta}$.  The estimate $\hat{\beta}^{(g+1)}$ in the $(g+1)$th iteration can be solved based on the previous coefficient vector $\hat{\beta}^{(g)}$ in the $g$th iteration:
\begin{equation*}
	\hat{\beta}^{(g+1)}=\hat{\beta}^{(g)}+[V^{(g)}+nH^{(g)})]^{-1}[P^{(g)}-nH^{(g)}\hat{\beta}^{(g)}],
\end{equation*}
with $P^{(g)}$ and $V^{(g)}$ as the first and second order derivative of the score function of QIF, respectively. They are defined as:
\begin{equation*}
	P^{(g)}=\frac{\partial Q(\hat{\beta}^{(g)})}{\partial \beta}=2\frac{\partial \overbar{\phi_n}^\top}{\partial \beta} \overbar{\Omega_n}^{-1}\overbar{\phi_n}(\hat{\beta}^{(g)}),
\end{equation*}
\begin{equation*}
	V^{(g)}=\frac{\partial^2 Q(\hat{\beta}^{(g)})}{\partial^2 \beta}=2\frac{\partial \overbar{\phi_n}^\top}{\partial \beta} \overbar{\Omega_n}^{-1}\frac{\partial \overbar{\phi_n}}{\partial \beta}.
\end{equation*}
$H^{(g)}$ is a diagonal matrix containing the derivatives of the penalty function and it's defined as:
\begin{equation*}
	\begin{aligned}
		H^{(g)}&=\text{diag}(\underbrace{0,...,0}_{1+q},\underbrace{\frac{\rho'(||\hat{\eta}_{1}^{(g)}||_{\Sigma_{1}};\sqrt{q+1}\lambda_{1},\gamma)}{\epsilon+||\hat{\eta}_{1}^{(g)}||_{\Sigma_{1}}},...,\frac{\rho'(||\hat{\eta}_{1}^{(g)}||_{\Sigma_{1}};\sqrt{q+1}\lambda_{1},\gamma)}{\epsilon+||\hat{\eta}_{1}^{(g)}||_{\Sigma_{1}}}}_{1+q},...,\\ &\underbrace{\frac{\rho'(||\hat{\eta}_{p}^{(g)}||_{\Sigma_{p}};\sqrt{q+1}\lambda_{1},\gamma)}{\epsilon+||\hat{\eta}_{p}^{(g)}||_{\Sigma_{p}}},...,\frac{\rho'(||\hat{\eta}_{p}^{(g)}||_{\Sigma_{p}};\sqrt{q+1}\lambda_{1},\gamma)}{\epsilon+||\hat{\eta}_{p}^{(g)}||_{\Sigma_{p}}}}_{1+q}),
	\end{aligned}
\end{equation*}
where $\lambda_1$ is the tuning parameter of genetic effects and gene-environment interactions and $\gamma$ is the regularization parameter. The first $(1 + q)$ elements on the diagonal of matrix $H$ are set to zero, since there is no shrinkage imposed on the intercept and the coefficients of the environmental factors. We can use $nH\hat{\beta}$ and $nH$ to approximate the first and second drivative functions of the the group MCP penalty. Starting with an inital coefficient vector,  we can repeat the proposed algorithm and update the regression parameter $\hat{\beta}^{(g+1)}$ through iterations.  We set the stop criterion $\text{mean}(|\hat{\beta}^{(g+1)}-\hat{\beta}^{(g)}|)<0.001$ and convergence can usually be achieved in a small to moderate number of iterations.

\subsection{Penalized QIF}

iQIF.exch, iQIF.ar1 and iQIF.ind are the second category of alternative methods considering only the individual level effects. The derivations for the three methods proceeds in a similar fashion. We have the penalized score function as:

\begin{equation*}
	U(\beta)=Q(\beta)+\sum_{v=1}^{p} \sum_{u=1}^{q+1}\rho  (|\eta_{vu}|;\lambda_{1},\gamma),
\end{equation*}
where  $\eta_{vu}$ denotes the $u$th element of $\eta_v$.  The Newton-Raphson update of $\hat{\beta}$ can be obtained as: 
\begin{equation*}
	\hat{\beta}^{(g+1)}=\hat{\beta}^{(g)}+[V^{(g)}+nH^{(g)})]^{-1}[P^{(g)}-nH^{(g)}\hat{\beta}^{(g)}],
\end{equation*}
where $P^{(g)}$ and $V^{(g)}$ are given as the corresponding first and second order derivatives of the score function of QIF as follows:
\begin{equation*}
	P^{(g)}=\frac{\partial Q(\hat{\beta}^{(g)})}{\partial \beta}=2\frac{\partial \overbar{\phi_n}^\top}{\partial \beta} \overbar{\Omega_n}^{-1}\overbar{\phi_n}(\hat{\beta}^{(g)}),
\end{equation*}
\begin{equation*}
	V^{(g)}=\frac{\partial^2 Q(\hat{\beta}^{(g)})}{\partial^2 \beta}=2\frac{\partial \overbar{\phi_n}^\top}{\partial \beta} \overbar{\Omega_n}^{-1}\frac{\partial \overbar{\phi_n}}{\partial \beta}.
\end{equation*}
The main diagonal of the diagonal matrix $H^{(g)}$ consists of the first order derivative of MCP:
\begin{equation*}
	\begin{aligned}
		H^{(g)}&=\text{diag}(\underbrace{0,...,0}_{1+q}, \underbrace{\frac{\rho'(|\hat{\eta}_{11}^{(g)}|;\lambda_{2},\gamma)}{\epsilon+|\hat{\eta}_{11}^{(g)}|},...,\frac{\rho'(|\hat{\eta}_{1(q+1)}^{(g)}|;\lambda_{2},\gamma)}{\epsilon+|\hat{\eta}_{1(q+1)}^{(g)}|}}_{1+q},...,\\ &\underbrace{\frac{\rho'(|\hat{\eta}_{p1}^{(g)}|;\lambda_{2},\gamma)}{\epsilon+|\hat{\eta}_{p1}^{(g)}|},...,\frac{\rho'(|\hat{\eta}_{p(q+1)}^{(g)}|;\lambda_{2},\gamma)}{\epsilon+|\hat{\eta}_{p(q+1)}^{(g)}|}}_{1+q}),
	\end{aligned}
\end{equation*}
where $\lambda_2$ and $\gamma$ are the tuning and regularization parameters, respectively. There is no shrinkage on the intercept and the coefficients of the environmental factors.  Hence the first $(1 + q)$ elements on the diagonal of matrix $H$ are set to zero. Here $nH\hat{\beta}$ and $nH$ can also be used to approximate the first and second drivative functions of the the MCP penalty. The iterative update of  $\hat{\beta}$ can be conducted till convergence.

\section{Other Simulation Results}
\label{appendix:c}
\setcounter{table}{0}
\renewcommand{\thetable}{B\arabic{table}}
\begin{table} [h!]
	\begin{center}
		\caption{Identification results for Scenario 3. TP/FP: true/false positives.  mean(sd) of  TP and FP based on 100 replicates.}\label{tab:3}
		
		\begin{tabular}{ l c c c c c c }
			\hline
			&\multicolumn{2}{c}{Overall }&\multicolumn{2}{c}{Main } &\multicolumn{2}{c}{Interaction } \\
			\cline{2-7}
			&TP&FP&TP&FP&TP&FP\\
			\hline
			sgQIF.exch&19.4(1.0)&2.1(1.1)&3.1(1.1)&0.9(0.7)&16.3(0.8)&1.3(0.8)\\
			
			gQIF.exch&22.1(1.6)&19.6(6.0)&4.9(0.9)&1.1(1.1)&17.3(1.0)&18.4(5.2)\\
			
			iQIF.exch&19.7(1.4)&9.0(4.8)&3.1(1.2)&2.0(1.2)&16.6(1.1)&7.0(4.0)\\
			
			\hline
			
			sgQIF.ar1&19.6(1.3)&2.8(1.3)&3.2(1)&0.6(0.7)&16.5(0.8)&2.2(1.4)\\
			
			gQIF.ar1&22.1(1.4)&18.5(5.3)&4.6(0.9)&0.9(1)&17.5(0.8)&17.6(4.5)\\
			
			iQIF.ar1&20.0(1.5)&9.2(4.4)&3.5(1.3)&1.6(1.3)&16.5(0.9)&7.5(3.6)\\
			
			\hline
			
			sgQIF.ind&19.3(1.5)&3.0(2.6)&3.0(1.0)&0.3(0.6)&16.3(1.5)&2.7(2.1)\\
			
			gQIF.ind&21.7(1.2)&16.3(5.1)&4.3(1.2)&1.7(1.2)&17.3(1.2)&14.7(4.0)\\
			
			iQIF.ind&20.0(1.0)&8.0(4.4)&3.0(1.0)&1.0(1.0)&17.0(1.0)&7.0(3.5)\\
			\hline 
		\end{tabular}
	\end{center}
	\centering
\end{table}

\begin{table} [h!]
	\begin{center}
		\caption{Identification results for Scenario 4. TP/FP: true/false positives.  mean(sd) of  TP and FP based on 100 replicates.}\label{tab:4}
		
		\begin{tabular}{ l c c c c c c }
			\hline
			&\multicolumn{2}{c}{Overall }&\multicolumn{2}{c}{Main } &\multicolumn{2}{c}{Interaction } \\
			\cline{2-7}
			&TP&FP&TP&FP&TP&FP\\
			\hline
			sgQIF.exch&21.9(1.6)&4.7(2.4)&6.6(0.5)&0.2(0.1)&15.3(1.5)&4.5(2.4)\\
			
			gQIF.exch&21.1(2.7)&19.2(4.4)&6.4(0.9)&3.1(1.6)&14.7(2.0)&16.1(3.1)\\
			
			iQIF.exch&22.3(1.3)&7.1(2.3)&6.8(0.5)&0.1(0.1)&15.5(1.2)&7.1(2.3)\\
			
			\hline
			
			sgQIF.ar1&22.3(2.1)&4.0(1.0)&6.7(0.6)&0.1(0.1)&15.7(1.5)&4.0(1.0)\\
			
			gQIF.ar1&22.4(1.9)&17.1(7.4)&6.9(0.4)&2.4(2.2)&15.6(1.8)&14.7(5.5)\\
			
			iQIF.ar1&23.3(0.6)&10.0(3.5)&7.0(0.6)&0.3(0.1)&16.3(0.6)&10.0(3.5)\\
			
			\hline
			
			sgQIF.ind&20.3(1.0)&3.5(1.3)&5.8(0.5)&0.1(0.1)&14.5(0.6)&3.5(1.3)\\
			
			gQIF.ind&22.5(0.7)&16.5(6.2)&7(0.3)&2.5(2.1)&15.5(0.7)&14.0(4.0)\\
			
			iQIF.ind&21.8(0.5)&9.5(1.3)&6.8(0.5)&0.1(0.1)&15.0(0.8)&9.5(1.3)\\
			\hline 
		\end{tabular}
	\end{center}
	\centering
\end{table}

\setcounter{figure}{0}
\renewcommand{\thefigure}{B\arabic{figure}}

\begin{figure}[H]
	\caption{Prediction (MSE) results of the 4 scenarios. mean(sd) of  prediction error based on 100 replicates.}
	\includegraphics[width=1\textwidth]{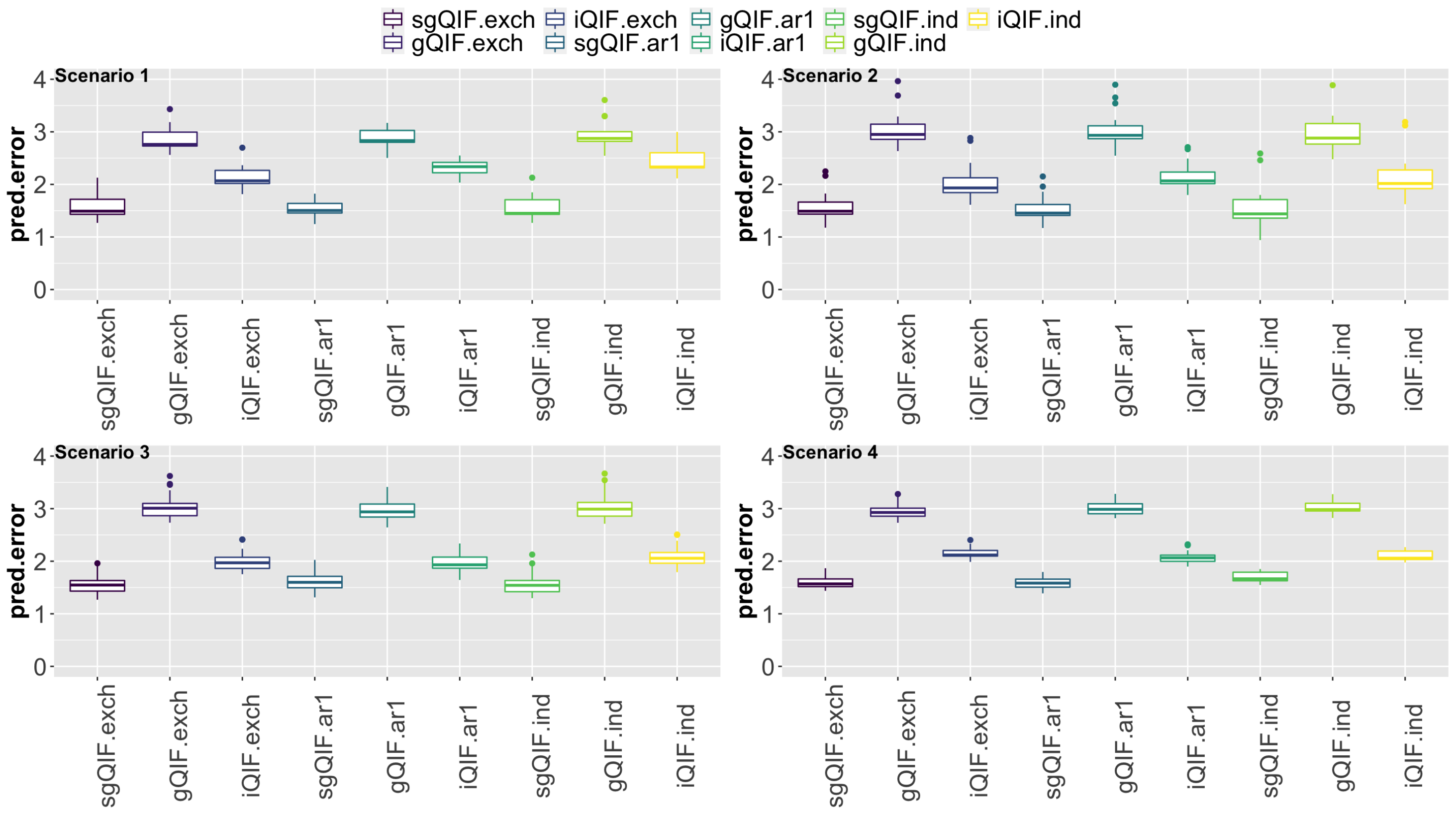}
	\label{fig:2}
\end{figure}

\section{Real Data Analysis}
\label{appendix:a}
\setcounter{table}{0}
\renewcommand{\thetable}{C\arabic{table}}
\begin{longtable} [ht!]{c c c c c c}
	\caption{Identification results on CAMP data using the bi-level selection method under the exchangeable working correlation (sgQIF.exch). The identified SNPs and the corresponding genes are listed in the first two columns.  The third column contains the coefficients of the main effects for each SNP.  The last three columns correspond to the interactions between the SNPs and environmental factors.}\label{tab:5}\\
	
	\hline
	SNP&Gene&&trt&age&gender\\
	
	\hline
	rs1276888&FAM46A&0&0.116&0&0\\
	
	rs10139964&AKAP6&0&0&0.125&0\\
	
	rs10852830&AC005703.2&0&0&0.111&0\\
	
	rs10995722&RP11-170M17.1&-0.398&0.103&0&0\\
	
	rs329614&NDUFAF2&0&0&0&0.598\\
	
	rs17431749&DKK2&0&-0.246&0&0.325\\
	
	rs2453021&TNFRSF9&0&0&0&-0.123\\
	
	rs1922134&RP11-170M17.1&-0.143&0&0&0\\
	
	rs290505&NDUFAF2&-0.155&0&0&0.300\\
	
	rs4969059&SLC39A11&-0.212&0&0&0\\
	
	rs4730738&CAV2&0&0&0&0.145\\
	
	rs162240&NDUFAF2&0.198&0&0&0\\
	
	rs6869332&ELOVL7&-0.246&-0.221&0&0\\
	
	rs167912&NDUFAF2&0&0&-0.282&0\\
	
	rs158928&ERCC8&0&-0.151&0&0.214\\
	
	rs131815&NCAPH2&0&0&0.129&0\\
	
	rs4280657&AC144521.1&0&0&0&-0.274\\
	
	rs11778333&TOX&0&0&-0.299&0\\
	
	rs11803207&KCND3&0&0&0&-0.139\\
	
	rs12299421&rs12299421&0&0&-0.105&0\\
	
	rs11257102&PFKFB3&-0.582&0&0&-0.353\\
	
	rs8141896&MICAL3&0.218&0&0&-0.152\\
	
	rs162231&NDUFAF2&0&-0.347&0&0\\
	
	rs10857493&RP11-123B3.2&0.468&-0.123&-0.495&0\\
	
	rs11257103&PFKFB3&-0.508&0&0&-0.192\\
	
	rs1251577&ST6GALNAC3&0&0&0.112&0\\
	
	rs4897284&LAMA2&0&0&0&-0.177\\
	
	rs10491881&RP11-202G18.1&0.128&-0.342&0.339&0\\
	
	rs566979&CAT&0&0&-0.105&0\\
	
	rs4904516&FOXN3&0&0.178&0&0\\
	
	rs681561&PCCA&0&0&-0.286&0.258\\
	
	rs1618870&CATSPERB&0&0.432&0&-0.489\\
	
	rs17010079&RP11-123B3.2&-0.214&0&0.364&0\\
	
	rs11031570&RCN1&0.291&0&-0.504&0\\
	
	rs909768&RPS6KA2&0&0&0&0.304\\
	
	rs9891809&SLC39A11&0&0.158&0&0\\
	
	rs8079240&SLC39A11&0&0.158&0&0\\
	
	rs7951816&SYT9&0&0&0.141&0\\
	
	rs1180286&CAV2&0&0&0&-0.152\\
	
	rs17813724&RP11-202G18.1&0&0.209&0&0.192\\
	
	rs17241424&TOX&-0.147&0&0.270&0\\
	
	rs11708933&AC144521.1&0&0&0&0.423\\
	
	rs197394&FAM212B&0&0&-0.276&0\\
	
	rs6008813&CELSR1&0&0.142&0&-0.119\\
	
	rs742267&RPS6KA2&0&0&-0.194&0\\
	
	rs7712473&ELOVL7&0&0&0&-0.154\\
	
	rs1704630&CATSPERB&0&0.638&0&-0.493\\
	
	rs10995701&RP11-170M17.1&0&0&-0.312&0\\
	
	rs4647078&ERCC8&-0.105&0.515&0&-0.115\\
	
	rs6877849&ELOVL7&0&0&0.405&0\\
	
	rs7029556&RP11-63P12.6&0&-0.119&0&0\\
	
	rs6449502&ELOVL7&0&0&-0.266&0\\
	
	rs12101359&UNC13C&0.107&0&0&0\\
	
	rs4716370&RP1-137D17.1&-0.227&0&0.215&0\\
	
	rs12060403&SLC35F3&0&0&-0.139&0\\
	
	rs12071173&SLC35F3&0&0&-0.139&0\\
	
	rs513555&SPRR2G&0&-0.289&0.291&0\\
	
	rs767006&CYFIP2&0.198&0&0&-0.164\\
	
	rs4700398&ELOVL7&0&0.205&0&-0.480\\
	
	rs197380&FAM212B&0.254&0&-0.479&0\\
	
	rs6914953&F13A1&-0.318&0&0&-0.227\\
	
	rs264356&NRG2&0&0&0.189&0\\
	
	rs10972815&CLTA&0&-0.108&0&0.253\\
	
	rs4700392&ELOVL7&-0.279&-0.916&0&0\\
	
	rs13194966&F13A1&-0.426&0&0.664&0\\
	
	rs1119266&SPRR2B&0&0.186&0&0\\
	
	rs11031563&RCN1&0.423&0&-0.549&0\\
	
	rs12101884&UNC13C&-0.225&0&0.436&0\\
	
	rs4647108&ERCC8&0.239&0&0&-0.607\\
	
	rs7718320&IQGAP2&-0.192&0&0.561&0\\
	
	rs2303921&TAF1B&0&0&0&-0.174\\
	
	rs1136062&CCNF&0&-0.125&0.191&0\\
	
	rs17390967&SCARA5&0.107&-0.196&0&-0.101\\
	
	rs7243734&ZBTB7C&-0.544&0&0&0\\
	
	rs17023415&AFF3&-0.305&0.383&0&0\\
	
	rs10995687&RP11-170M17.1&0.169&0&-0.294&0\\
	
	rs13265701&MYOM2&-0.218&0.263&0&0\\
	
	rs4940195&ZBTB7C&-0.51&0&0&0\\
	
	rs2918528&ZNF717&0&0&0.290&-0.148\\
	
	rs17819589&RP11-392P7.6&0&0&-0.157&0\\
	
	rs1360176&RP11-82L2.1&-0.268&0&0&0.213\\
	
	rs17660456&MYO5B&-0.138&0.157&0&0\\
	
	rs10871386&RP11-525K10.3&0&0&-0.105&0.201\\
	\hline 
	\centering
\end{longtable}

\begin{longtable} [ht!]{c c c c c c}
	\caption{Identification results on CAMP data using the individual-level selection method under the exchangeable working correlation (iQIF.exch). The identified SNPs and the corresponding genes are listed in the first two columns.  The third column contains the coefficients of the main effects for each SNP.  The last three columns correspond to the interactions between the SNPs and environmental factors.}\label{tab:6}\\
	
	\hline
	SNP&Gene&&trt&age&gender\\
	
	\hline
	rs10050758&SLC36A2&0&-0.136&0.135&0\\
	
	rs1276888&FAM46A&0&0&0&-0.168\\
	
	rs10852830&AC005703.2&0&-0.138&0&0\\
	
	rs10995722&RP11-170M17.1&-0.419&0&0.413&0\\
	
	rs329614&NDUFAF2&0&0.153&0&0\\
	
	rs1922134&RP11-170M17.1&0.178&0&0.332&0\\
	
	rs290505&NDUFAF2&0&0&0&0.429\\
	
	rs4969059&SLC39A11&0&0.257&0&0\\
	
	rs4730738&CAV2&0&0&-0.132&0\\
	
	rs162240&NDUFAF2&-0.374&0&0&0.817\\
	
	rs6869332&ELOVL7&-0.498&0.696&0&0\\
	
	rs167912&NDUFAF2&-0.210&0&0&0.378\\
	
	rs158928&ERCC8&0&0.314&0&0\\
	
	rs131815&NCAPH2&0&0&0.146&0\\
	
	rs4280657&AC144521.1&0&0.222&-0.250&0\\
	
	rs11778333&TOX&0&-0.235&0&0.255\\
	
	rs11803207&KCND3&0&0&-0.241&0.175\\
	
	rs11257102&PFKFB3&-0.238&0&0&-0.233\\
	
	rs8141896&MICAL3&0.278&0&-0.370&0\\
	
	rs162231&NDUFAF2&0&-0.262&0&-0.218\\
	
	rs10857493&RP11-123B3.2&0.451&0&-0.536&0\\
	
	rs10796011&CCDC3&0&0&0&0.145\\
	
	rs11257103&PFKFB3&-0.170&0&0&0\\
	
	rs1251577&ST6GALNAC3&0&0&0.133&0\\
	
	rs4897284&LAMA2&0&-0.273&0.290&0\\
	
	rs10491881&RP11-202G18.1&0&-0.169&0&0\\
	
	rs4904516&FOXN3&0&0.247&0&0\\
	
	rs681561&PCCA&0&0&-0.258&0.263\\
	
	rs1618870&CATSPERB&0&-0.16&0.686&-0.551\\
	
	rs17010079&RP11-123B3.2&-0.227&0&0.428&0\\
	
	rs11031570&RCN1&0.410&-0.670&0&0\\
	
	rs909768&RPS6KA2&0&0&0&0.142\\
	
	rs7951816&SYT9&0&0&0.133&0\\
	
	rs1180286&CAV2&0&0&0&-0.221\\
	
	rs17241424&TOX&0&0&0&0.210\\
	
	rs17044664&AC144521.1&0&0.248&0&-0.159\\
	
	rs11708933&AC144521.1&0&0&0&0.163\\
	
	rs197394&FAM212B&0&0.270&0&-0.224\\
	
	rs6008813&CELSR1&0&0.156&0&0\\
	
	rs742267&RPS6KA2&0&0&-0.371&0\\
	
	rs742269&RPS6KA2&0&-0.144&0&0\\
	
	rs7712473&ELOVL7&-0.271&-0.567&0&0.722\\
	
	rs1704630&CATSPERB&0&0&0.569&-0.575\\
	
	rs17015079&ROBO2&-0.357&0.548&0&0\\
	
	rs10995701&RP11-170M17.1&0&0&0&-0.348\\
	
	rs4647078&ERCC8&-0.214&0&0&0.335\\
	
	rs6877849&ELOVL7&0&0&0.993&-0.414\\
	
	rs7029556&RP11-63P12.6&0&0&0.315&0\\
	
	rs6449502&ELOVL7&0&0&0.491&-0.867\\
	
	rs12101359&UNC13C&0.133&0&0&0\\
	
	rs34673&TNPO1&0&0.198&0&0\\
	
	rs12060403&SLC35F3&0&0&-0.344&0\\
	
	rs12071173&SLC35F3&0&0&-0.344&0\\
	
	rs12073596&SLC35F3&0&0&-0.159&0\\
	
	rs12085211&SLC35F3&0&0&-0.159&0\\
	
	rs1545854&LINC00880&0&0.139&0&-0.149\\
	
	rs513555&SPRR2G&0&-0.290&0&0\\
	
	rs4700398&ELOVL7&-0.507&0&0&0.160\\
	
	rs197380&FAM212B&0.224&0&-0.273&0\\
	
	rs6914953&F13A1&0&0&0&-0.275\\
	
	rs264356&NRG2&0&0&0.270&0\\
	
	rs463221&CTD-2193G5.1&-0.156&0.154&0&0\\
	
	rs4700392&ELOVL7&0.365&-0.982&0&0\\
	
	rs13194966&F13A1&-0.178&0&0.391&0\\
	
	rs1119266&SPRR2B&0&0.291&-0.264&0\\
	
	rs11031563&RCN1&0.542&-0.474&0&0\\
	
	rs12101884&UNC13C&-0.176&0&0.364&0\\
	
	rs4647108&ERCC8&0&0.319&0&-0.389\\
	
	rs719628&TASP1&0&0.154&0&0\\
	
	rs7718320&IQGAP2&0&0&0.355&-0.409\\
	
	rs1136062&CCNF&0&-0.131&0.135&0\\
	
	rs7243734&ZBTB7C&-0.703&0&0&0\\
	
	rs17023415&AFF3&-0.225&0.353&0&0\\
	
	rs17128269&SH2D4A&0&0&-0.342&0.196\\
	
	rs10995687&RP11-170M17.1&0&0&-0.173&0.171\\
	
	rs10734883&SLC2A14&-0.133&0.173&0&0\\
	
	rs13265701&MYOM2&-0.245&0.157&0&0\\
	
	rs4940195&ZBTB7C&-0.708&0&0&0\\
	
	rs2918528&ZNF717&0&0&0.218&0\\
	
	rs1360176&RP11-82L2.1&0&0&0&0.246\\
	
	rs17660456&MYO5B&0.229&0&-0.398&0\\
	
	rs10871386&RP11-525K10.3&0&0&-0.199&0.184\\
	\hline 
	\centering
\end{longtable}

\end{document}